\documentclass[pra,twocolumn,amsmath,amssymb,superscriptaddress,eqsecnum]{revtex4-2}
\usepackage{graphicx,amsmath,relsize,epstopdf,color,mathtools,bm,newtxtext,newtxmath,braket,rotating}
\usepackage[hyphenbreaks]{breakurl}
\usepackage[colorlinks=true,linkcolor=blue,citecolor=blue,urlcolor =blue]{hyperref}
\usepackage{soul}

\usepackage[table,xcdraw]{xcolor}
\newcommand{\Tr}{\mathop{\mathrm{Tr}} \nolimits}

\begin{document}
\setstcolor{red}

\title{Multipoles from Majorana constellations}

\author{J. L. Romero}
\affiliation{Departamento de  \'Optica, Facultad de F\'{\i}sica, Universidad Complutense, 28040~Madrid, Spain}

\author{A. B. Klimov}
\affiliation{Departamento de F\'{\i}sica, Universidad de Guadalajara, 44420~Guadalajara, Jalisco, Mexico}

\author{A. Z. Goldberg}
\affiliation{National Research Council of Canada, Ottawa, Ontario K1N 5A2, Canada}
\affiliation{Department of Physics, University of Ottawa, Ottawa, Ontario K1N 6N5, Canada}

\author{G. Leuchs}
\affiliation{Max-Planck-Institut f\"ur die Physik des Lichts, 91058 Erlangen, Germany}
\affiliation{Institut f\"{u}r Optik, Information und Photonik, Friedrich-Alexander-Universit\"{a}t Erlangen-N\"{u}rnberg, 91058~Erlangen, Germany}

\author{L. L. S\'anchez-Soto}
\affiliation{Departamento de  \'Optica, Facultad de F\'{\i}sica, Universidad Complutense, 28040~Madrid, Spain}
\affiliation{Max-Planck-Institut f\"ur die Physik des Lichts, 91058 Erlangen, Germany}

\begin{abstract}
Majorana stars, the $2S$ spin coherent states that are orthogonal to a spin-$S$ state, offer an elegant method to visualize quantum states,  disclosing their intrinsic symmetries. These states are naturally described by the corresponding multipoles. These quantities can be experimentally determined and allow for an SU(2)-invariant analysis.  We investigate the relationship between Majorana constellations and state multipoles, thus providing insights into the underlying symmetries of the system.  We illustrate our approach with some relevant and informative examples.
\end{abstract}

\maketitle
\section{Introduction}

Completely symmetric states aptly describe numerous phenomena~\cite{Eckert:2002aa}. These states play a crucial role in the characterization of spinor Bose gases~\cite{Kawaguchi:2012aa,Stamper-Kurn:2013aa}, whose dynamics have been observed  with spin-1~\cite{Anderson:1995aa,Davis:1995aa}, spin-2~\cite{Chang:2004aa}, and even spin-3~\cite{Griesmaier:2005aa,Beaufils:2008aa} condensates.  They have also been used for the characterization of entanglement~\cite{Devi:2007wy,Bastin:2009ta,Martin:2010ui,Aulbach:2012wb,Mandilara:2014aa,Goldberg:2022aa} in those boson systems.

A salient feature of this family is that any state of spin $S$ can be described as a {permutation-symmetric} $n$-qubit state, with $n=2S$.  This is the basis of the elegant representation devised by Majorana~\cite{Majorana:1932ul}, in which a spin-$S$ is depicted by $2S$ points (called the stars) on the Bloch sphere. Although the subject attracted some attention in relation to the quantum theory of angular momentum~\cite{Bloch:1945aa,Salwen:1955aa,Meckler:1958aa,Schwinger:1977aa}, it remained quiescent until 1969, when Penrose reinterpreted Majorana stars as principal null directions in spinor theory~\cite{Penrose:1960aa} and brought it to wider attention in his celebrated book~\cite{Penrose:1989aa}. 

Apart from indisputable mathematical advantages~\cite{Bacry:2004aa}, this picture builds a bridge between the abstract Hilbert space (where the states live) and the simple geometry of the Bloch sphere. Consequently, this representation rapidly meets the increasing interest in high-dimensional quantum systems and, several decades after its conception, is being used in fields as diverse as polarization~\cite{Hannay:1998ab,Bjork:2015aa,Goldberg:2019aa,Goldberg:2021tx,Goldberg:2022ab}, spinor Bose gases~\cite{Barnett:2009aa,Lamacraft:2010aa,Lian:2012aa,Cui:2013aa},  multiqubit systems~\cite{Makela:2010aa,Ribeiro:2011ti,Devi:2012aa,Chryssomalakos:2021aa,Kam:2020aa,Lu:2021aa,Burchardt:2021aa}, metrology~\cite{Bouchard:2017aa,Chryssomalakos:2017aa,Goldberg:2018aa,Martin:2020aa,Goldberg:2021aa}, geometric phases~\cite{Hannay:1998aa,Bruno:2012aa,Liu:2014aa,Yang:2015aa,Liu:2016aa,Yao:2017aa,Kam:2021aa,Mittal:2022aa}, non-Hermitian lattices~\cite{Bartlett:2021aa} and algebraic quantum models, such as the Lipkin-Meshkov-Glick model~\cite{Ribeiro:2007aa,Ribeiro:2008aa}.

The distribution of Majorana stars conveys complete information  and can be directly computed when the quantum state is known. But this notion has far-reaching advantages: by visualizing in a crystal-clear manner the intrinsic symmetries of the state, one can perceive connections with other intriguing questions. Examples include geometrical measures of entanglement~\cite{Markham:2011wa,Wang:2012aa,Ganczarek:2012aa}, spherical $t$-designs~\cite{Delsarte:1977dn,Hardin:1992mi}, and the Thomson~\cite{Thomson:1904qp,Melnyk:1977gm,Ashby:1986bk,Erber:1991aa,Edmundson:1992uf,Baecklund:2014ng} and Tammes~\cite{Tammes:1930rc,Fejes:1953aa,Leech:1957aa} problems. Moreover, a number of states with remarkable properties, such as Queens~\cite{Giraud:2010db} and Kings of Quantumness~\cite{Bjork:2015ab}, maximally entangled states~\cite{Aulbach:2010ui}, $k$-uniform states~\cite{Arnaud:2013hm,Goyeneche:2014aa,Li:2019aa}, and states with maximal Wehrl entropy~\cite{Lieb:2014aa}, can be aptly understood in terms of the properties of their corresponding constellations~\cite{Goldberg:2020aa}. 

Distributing points on a sphere is a mathematical problem with a long history and with a variety of optimal configurations depending on the cost function one tries to optimize~\cite{Whyte:1952aa,Saff:1997aa,Katanforoush:2003aa,Brauchart:2015aa}.  This suggests exploring those arrangements of points distinguished by some extremal properties. Then a natural question arises: if one knows the locations of the Majorana stars, what can one say about the state, in particular, about its multipolar distribution? That is precisely our main goal here.

For a system of point charges on the sphere, the most suitable way of capturing the progressively finer angular features of the system is the standard multipolar expansion~\cite{Jackson:1999aa}. Such an expansion can often be truncated, meaning that to a good approximation only the first terms need to be retained. We propose here to carry out a similar procedure for the Majorana constellation: the resulting multipoles constitute a basic tool for problems with an SU(2) invariance~\cite{Fano:1959ly,Blum:1981rb} and, in addition, they can be experimentally determined with simple procedures~\cite{Bjork:2012zr,Goldberg:2022ac}. 

The paper is organized as follows. In Sec.~\ref{sec:const} we introduce the basic notions needed to understand the Majorana constellations, whereas in Sec.~\ref{sec:multi} we show how to calculate the multipoles from a given Majorana constellations. Our insight is to apply the time-honored Vieta formulas~\cite{Funkhouser:1930aa} to the polynomial defining the constellation, which provides a shortcut between the stars and other representations of the state. In Sec.~\ref{sec:exam} we thoroughly examine how the method works in a series of relevant examples.  Further physical implications are discussed in Sec.~\ref{sec:fur} and, finally, our conclusions are summarized in Sec.~\ref{sec:conc}.

\section{Majorana constellations}
\label{sec:const}

We will deal with any pure system living in a finite-dimensional Hilbert space of dimension $2S+1$, which can be formally regarded as a spin-$S$. The corresponding space $\mathcal{H}_{S}$, spanned by the standard angular momentum basis  $\{ |S, m\rangle \mid m= -S, \ldots, S\}$, is the carrier of the irreducible representation (irrep) of spin $S$ of SU(2) and is isomorphic to $\mathbb{C}^{2S+1}$. Since any two vectors in $\mathcal{H}_{S}$ differing by a phase represent the same physical state,  the manifold of physical states is the projective space $\mathbb{C}\mathbf{P}^{2S}$~\cite{Bengtsson:2006aa}. 

The merit of Majorana was to show that points in $\mathbb{C}\mathbf{P}^{2S}$ are in one-to-one correspondence with unordered sets of (possibly coincident) $2S$ points on the unit sphere $\mathcal{S}_{2}$. There are various ways to see why this is so, but probably the most direct one is in terms of coherent states.

The spin (or Bloch) coherent states live in $\mathcal{H}_{S}$ and are displaced versions of a fiducial state, much the same as for the canonical coherent states on the plane. This fiducial state is chosen so as to minimize the variance of the Casimir operator $\mathbf{S}^{2} = S_{x}^{2} + S_{y}^{2} + S_{z}^{2}$, where $(S_{x}, S_{y}, S_{z})$ are the angular momentum operators, which generate the algebra $\mathfrak{su}(2)$. The minimum-variance states are $|S, \pm S \rangle$ and they guarantee that their displaced versions are the closest to classical states. The displacement operator on $\mathcal{S}_{2}$ is $D(\theta, \phi) = \exp(i \phi S_{z}) \exp(i \theta S_{y}) = \exp [ \tfrac{1}{2} \theta (S_{+} e^{-i \phi} - S_{-} e^{i \phi})]$, where $S_{\pm} = S_{x} \pm i S_{y}$ are raising and lowering operators. Disentangling this  displacement allows us to express the coherent states $|\theta, \phi \rangle = D(\theta, \phi) |S, -S\rangle$   as~\cite{Perelomov:1986aa,Gazeau:2009aa}
\begin{equation}
|\theta, \phi \rangle \equiv |z \rangle = \frac{1}{( 1 + \lvert z \rvert^{2})^{S}} 
\exp (z {S}_{+}) \ket{S, -S} \, ,
\end{equation}
where the label  $z = \tan (\theta/2) e^{- i \phi}$ corresponds to an inverse stereographic projection from the south pole, mapping the point $z \in \mathbb{C}$ onto the point $(\theta, \phi) \in \mathcal{S}_{2}$~\cite{Coxeter:1969aa}. 

On expanding the exponential, we can write the coherent states in terms of the basis states of the irrep:
\begin{equation}
|z \rangle = \frac{1}{( 1 + \lvert z \rvert^{2})^{S}} \sum_{m=-S}^{S} 
\binom{2S}{S+m}^{\tfrac{1}{2}} \, z^{S+m} \ket{S, m} \, ,
\end{equation} 
or, employing again the stereographic projection  
\begin{align}
|z \rangle  & =  \sum_{m=-S}^{S} 
\binom{2S}{S+m}^{\tfrac{1}{2}}\; [\sin (\theta/2) ]^{S+m}   
[\cos (\theta/2) ]^{S-m} \nonumber \\
& e^{-i (S+m) \phi} \ket{S, m} \, .
\end{align} 

The system of spin coherent states is complete, but the states are not mutually orthogonal; their overlap is
\begin{equation}
\label{eq:overlap}
 \langle z | z^{\prime} \rangle =
 \frac{(1 + z^{\ast} z^{\prime})^{2S}}{\left [( 1 + \lvert z \rvert^{2}) ( 1 + \lvert z^{\prime} \rvert^{2}) \right ]^{S}} \, .
\end{equation} 
{They allow} for a resolution of the unity in the form
\begin{equation}
\int_{\mathbb{C}} d\mu_{S} (z) \, \ket{z} \bra{z} = \openone  \, ,
\end{equation}
with the invariant measure given by 
\begin{equation}
d\mu_{S} (z) = \frac{2S+1}{\pi} \frac{d^{2} z}{(1 + \lvert z \rvert^{2})^{2}} \, .
\end{equation} 
With this completeness relation one is able to decompose an arbitrary pure state over the coherent states.  If we denote $\psi (z^{\ast}) = \langle z | \psi \rangle$, by using the basis $\{ |S, m \rangle \}$, we define the stellar function $f_\psi (z)$ of the state $|\psi\rangle$ as  
\begin{align}
f_{\psi} (z) & = ( 1 + \lvert z \rvert^{2})^{S} \psi(z) = 
\sum_{m=-S}^{S} \binom{2S}{S+m}^{\tfrac{1}{2}} \psi_{m} \; z^{S+m} \nonumber \\
& = \sum_{k=0}^{2S} \binom{2S}{k}^{\tfrac{1}{2}} \psi_{k-S} \; z^{k}  \, ,
\end{align}
with $\psi_{m} = \langle S,m | \psi\rangle$ and in the second line we have made the relabeling $S+m \mapsto k$. Interestingly, in this representation the wave function is a polynomial in $z$ of  order $r \le 2S$. In consequence, the roots $z_{k} \in \mathbb{C}$ of ${f_{\psi} ( z)}$ fully characterize the state. These roots define, via an inverse stereographic map, $2S$ points on the unit sphere $\mathcal{S}_{2}$. This is the Majorana constellation, and each one of these points constitutes one star of the constellation.  Note that an SU(2) rotation  corresponds to a solid rotation of the constellation; therefore, states with the same constellation, irrespective of their relative orientation, have the same physical properties.

The stellar function is directly related to the Husimi $Q$ function~\cite{Husimi:1940aa,Kano:1965aa}:  
\begin{equation}
Q_\psi ( z ) =   ( 1 + \lvert z \rvert^{2})^{2S} \lvert f_{\psi} ( z^{\ast} ) \rvert^{2}  \, ,
\end{equation}
which clearly shows that the zeros of the Husimi $Q_\psi$ function are the complex conjugates of the zeros of $f_{\psi}$. These can then be observed in the laboratory by measuring where the Husimi function vanishes~\cite{Muller:2016aa}.

Let us examine a few relevant examples to illustrate how these constellations look. The first one is that of a spin coherent state $|z_{0} \rangle$, whose stellar representation is direct from Eq.~\eqref{eq:overlap}: 
\begin{equation}
f_{z_{0}} (z)  = \frac{(1 + z_{0}  z)^{2S}}
{(1 + | z_{0} |^{2})^{S}} 
\end{equation}
so it has  a single zero at $z= - 1/z_{0}$ with multiplicity $2S$. In consequence, the constellation collapses in this case to a single point diametrically opposed to the maximum $z_{0}$.

\begin{figure}
  \centering{\includegraphics[width=\columnwidth]{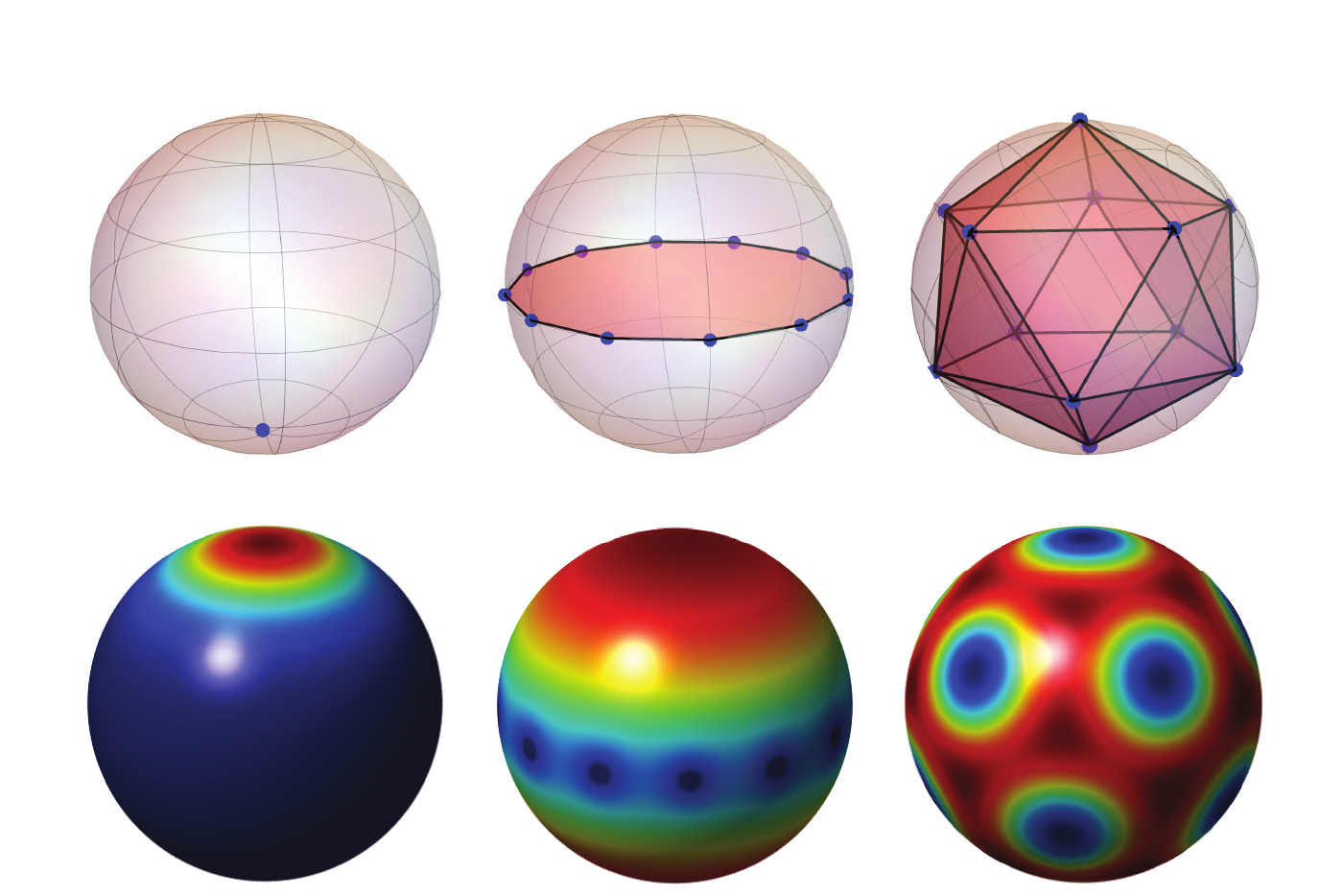}}
  \caption{{(Top) Majorana constellations for, from left to right, a spin coherent state, a  NOON state, and a  King of Quantumness, all of them for the same dimension with $S=6$. (Bottom) Density plots of the corresponding  Husimi functions, with a scale that goes from blue (minimum) to red (maximum).}}
  \label{Fig:Majorana}
\end{figure}

Another relevant set of states is that of so-called NOON states, defined as~\cite{Dowling:2008aa} 
\begin{equation}
 |\mathrm{NOON} \rangle =  \frac{1}{\sqrt{2}}
  {(|S,S\rangle - |S,-S\rangle)} \, .
  \label{Eq: NOON}
\end{equation}
They are known to have the highest sensitivity for a fixed excitation $S$ to small rotations about the $S_{z}$-axis~\cite{Bollinger:1996aa}. The associated polynomial reads
\begin{equation}
f_{\mathrm{NOON}} (z) = \frac{1}{\sqrt{2}}  (z^{2S} - 1)\, .
\end{equation}
The zeros are thus the $2S$ roots of unity, so the Majorana constellations have $2S$ stars placed around the equator  with equal angular separation between each star. A rotation around the $S_z$ axis of angle $\pi/(2 S)$ renders the state orthogonal to itself, justifying their optimality.

Finally, we consider Kings of Quantumness, initially dubbed anticoherent states~\cite{Zimba:2006fk}. In a sense they are the opposite of spin coherent states: whereas the latter correspond as nearly as possible to a classical spin vector pointing in a given direction, the former \emph{point nowhere}; i.e., the average angular momentum vanishes and the fluctuations up to given order $M$ are isotropic~\cite{Hoz:2013om}. Their symmetrical Majorana constellations herald their isotropic angular momentum properties~\cite{Markus:2015aa} {and correspond to the vertices of Platonic solids in some particular dimensions}.  In Fig.~\ref{Fig:Majorana}  we show the constellation associated to the examples aforementioned for the case of spin $S=6$.

\section{State multipoles from a constellation}
\label{sec:multi}

So far, we have shown how to compute the constellation when the state is given. In this section, we attack the inverse problem: what information can we extract from a given constellation. To this end, we first recall that every polynomial $P(z)$ of degree $n$ can be represented in terms of its zeros $\zeta_{k}$ using the classical Vieta formulas~\cite{Funkhouser:1930aa}, which can be expressed  in the form
	\begin{equation}
	\label{eq:polzeros1}
		P(z) = \sum_{k=0}^{n} a_{k} z^{k} = a_n \sum_{k=0}^{n} (-1)^{n-k} 
		e_{n-k}(\bm{\zeta} ) \; z^{k} \, ,
	\end{equation}
where $e_{j} (\bm{\zeta}) \equiv e_{j} (\zeta_{1}, \zeta_{2}, \ldots, \zeta_{n}) $ are the elementary symmetric polynomials~\cite{Waerden:1991aa} defined as
\begin{align}
e_{0} (\zeta_{1}, \zeta_{2}, \ldots, \zeta_{n}) & = 1 \, \nonumber \\
e_{1} (\zeta_{1}, \zeta_{2}, \ldots, \zeta_{n}) & = \sum_{1 \le j \le n} \zeta_{j} \, \nonumber \\
e_{2} (\zeta_{1}, \zeta_{2}, \ldots, \zeta_{n}) & = \sum_{1 \le j < k \le n} \zeta_{j} \zeta_{k} \,  \nonumber \\
& \vdots \nonumber \\
e_{n} (\zeta_{1}, \zeta_{2}, \ldots, \zeta_{n}) & = \zeta_{1} \zeta_{2} \ldots \zeta_{n} \, .
 \end{align} 
Using this fundamental result, the Majorana stellar function  can be expressed in the compact form   	
\begin{equation}
\label{eq:Majpolbo}
f_\psi (z ) = \sum_{k=0}^{2S} \mathfrak{f}_{k} (\bm{\zeta}) \; z^{k} \, ,
\end{equation}
where we have introduced the notation
\begin{equation}
\label{eq:defpsik}
		\mathfrak{f}_{k} (\bm{\zeta}) =  (-1)^{2S-k} \, \psi_{S} \, e_{2S-k}( \bm{\zeta}) \, ,
	\end{equation}
and the coefficient $\psi_{S}$ is fixed by the normalization condition
\begin{equation}
	\psi_{S}(\bm{\zeta})=\left(\sum_{k=0}^{2S}\frac{\left|{e_{2S-k}(\zeta)}\right|^2}{\binom{2S}{k}}\right)^{-1/2} \, .
\end{equation}
In this way, the state coefficients are simply related to the coefficients of the Majorana polynomial, and we can calculate them as a function of the stars:
\begin{equation}
\psi_{k-S}(\bm{\zeta})=\frac{\mathfrak{f}_{k}(\bm{\zeta})}{\sqrt{\binom{2S}{k}}} \, .
\end{equation}

For many purposes, expanding in the basis $|S,m\rangle$ is not a good choice. Instead, if one considers the associated density matrix $\varrho = | \psi \rangle \langle \psi |$,  it proves more convenient to use the  irreducible tensors~\cite{Blum:1981rb}. They are defined as  
\begin{equation}
  \label{Tensor} 
  {T}_{Kq} = \sqrt{\frac{2 K +1}{2 S +1}} 
  \sum_{m,  m^{\prime}= -S}^{S} C_{Sm, Kq}^{Sm^{\prime}} \, 
  |  S , m^\prime \rangle \langle S, m | \, ,
\end{equation}
with $ C_{Sm, Kq}^{Sm^{\prime}}$ being the Clebsch-Gordan coefficients that couple a spin $S$ and an integer-spin $K$ \mbox{($0 \le K \le 2S$)} to a total spin $S$~\cite{Varshalovich:1988xy}.  These tensors constitute  an orthonormal basis $\Tr  ({T}_{K q}\, {T}_{K^{\prime} q^{\prime}}^{\dagger} )  =  \delta_{K  K^{\prime}} \delta_{q q^{\prime}}$ and have the correct transformation properties under rotations.  

The corresponding expansion coefficients $\varrho_{Kq} =  \Tr ( {\varrho} \, {T}_{Kq}^{\dagger} )$ are known as state multipoles.  Actually,   $\varrho_{Kq}$ can be related to the $K$th powers of the generators. The monopole $\varrho_{00}= 1/\sqrt{2S+1}$ is trivially fixed by normalization; the dipole $\varrho_{1q}$ is the first-order moment of ${\mathbf{S}}$ and thus corresponds to the classical picture, in which the state is represented by its average value on the Bloch sphere. However, the complete characterization of a state demands the knowledge of the other multipoles that account for higher-order fluctuations.  The hermiticity of $\varrho$ imposes  the conditions $\varrho_{K-q} = ( -1)^{q} \, \varrho_{Kq}^\ast$.

The multipoles  can be expressed in terms of the state amplitudes $\psi_{m}$, which, in turn, can be computed from the constellation. This allows us to calculate $\varrho_{Kq}$ as a function of the constellation in a very compact form
\begin{align}
	\label{28}
	 \varrho_{Kq} & = \sqrt{\frac{2K+1}{2S+1}} \sum_{m=-S}^S C_{Sm,Kq}^{S,m+q} \, \psi_{m+q}^{\ast} ( \bm{\zeta} ) \, \psi_{m} (\bm{\zeta}) \, .
\end{align} 

It will prove convenient to characterize the multipoles by their effective length
\begin{equation}
\varrho_{K}^{2} = \sum_{q=-K}^{K} | \varrho_{Kq} |^{2} \, ,
\end{equation}
which gauges the state overlapping with the $K$th multipole pattern and is unchanged via SU(2) rotations of the state.  Note that, for pure states we are considering, $\sum_{K=0}^{2S} \varrho_{K}^{2} = 1$. These coefficients $\varrho_{K}^{2}$  have been used as measures of localization~\cite{Hall:1999oq}, as quantifiers of quantumness~\cite{Goldberg:2020aa}, which is useful for applications such as rotation sensing~\cite{Goldberg:2021aa}, and to quantify mode-decomposition-independent entanglement properties~\cite{Goldberg:2022aa}.

\begin{figure}
  \centering{\includegraphics[width=0.85\columnwidth]{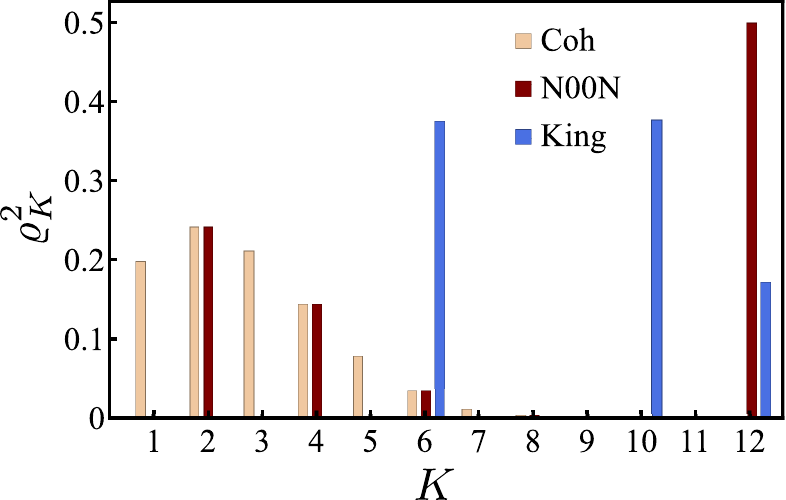}}
  \caption{Multipole lengths for the three constellations depicted in Fig.~\ref{Fig:Majorana}, as a function of the  order $K$.}
  \label{fig:multipoles}
\end{figure}

\section{Examples}
\label{sec:exam}

\begin{figure}[t]
  \centering{\includegraphics[width=0.90\columnwidth]{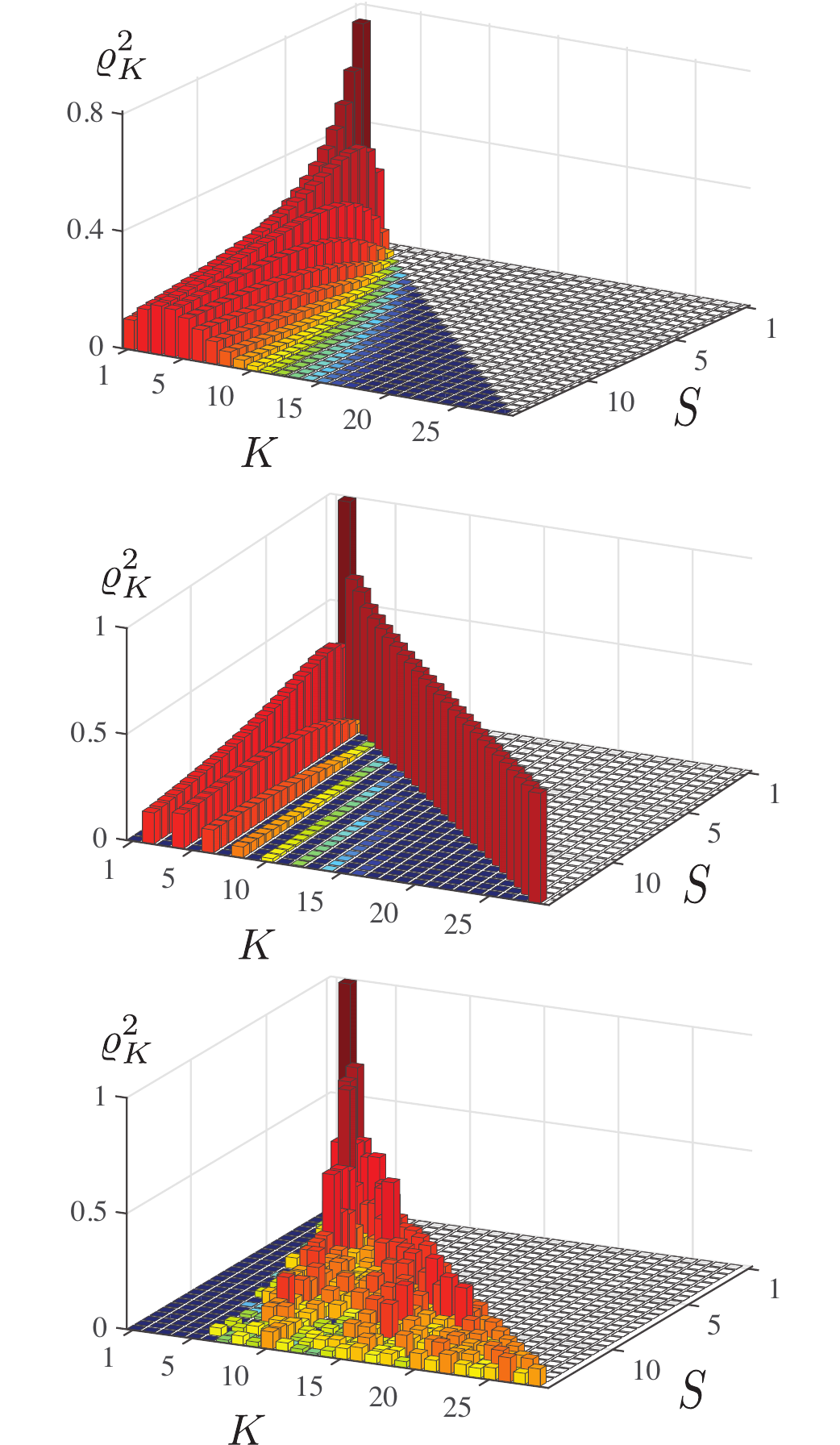}}
  \caption{Multipole lengths as a function of the order $K$ and the system spin $S$ for (upper panel) spin coherent states, (middle panel) NOON states, and (lower panel) Kings of Quantumness. Since the value of the monopole depends on the dimension, we normalized the rest of the multipole lengths to add up to 1. This allows us to compare their values for different $S$.} 
  \label{Fig:Multcoh}
\end{figure}

To check how the method works we consider the constellations depicted in Fig.~\ref{Fig:Majorana}, corresponding to the states worked out in Sec.~\ref{sec:const}; namely, coherent, NOON, and King of Quantumness for $S=6$. In Fig.~\ref{fig:multipoles} we represent the corresponding $\varrho_{K}^{2}$ as a function of the order $K$ (excluding the monopole, as it is always trivial). {Similar patterns emerge for other $S$.} For completeness, in Fig.~\ref{Fig:Multcoh}  we also plot  the distribution of $\varrho_{K}^{2}$ as a function of the order $K$ and the spin $S$ for the same three states.

In the Bloch sphere, constellations having their points arranged as symmetrically as possible are the most quantum, whereas the opposite occurs for coherent states. As we can appreciate, the coherent state conveys all the relevant information in the lowest-order multipoles, which is in agreement with its classical character.  Only for high values of $S$, one can see a tiny contribution of higher-order multipoles. In the limit of large $S$, we can fit a Gaussian distribution to these multipoles. Since the distributions are unchanged by rotations of the Bloch sphere, we can choose the coherent state with $z=0$ to find 
\begin{equation}
\varrho_{K, {\mathrm{coh}}}^2 = \frac{(2K+1)(2S)!^2}{(2S-K)!(2S+K+1)!} \, .
\end{equation}
The maximum of this distribution is at $K_{\mathrm{max}} =\sqrt{S+1/2}-1/2$. In the limit of large $S$, this peaks at $\varrho^2_{K_{\mathrm{max}}}\approx 1/\sqrt{Se}$, with a variance $S/2$.

The NOON state has only even multipoles which, surprisingly, are identical with the corresponding ones from the coherent state. We thus could regard the NOON state as a coherent state wherein the information from the odd multipoles has been transferred to the highest-order one.  In the limit of high $S$ only that multipole is relevant (see Fig.~\ref{Fig:Multcoh}), confirming that NOON states are more quantum for higher $S$. The significant contribution of the NOON state to the final multipole raises the question of the maximum value that a state can achieve in its highest-order multipole. To explore this question, we numerically generated $6 \times 10^{4}$ random constellations for various values of $S$:  none of them exhibited a larger highest-order multipole than the NOON state. This brings the question of the maximum value for the $K$th multipole length.

\begin{figure}[t]
	\centering{\includegraphics[width=0.95\columnwidth]{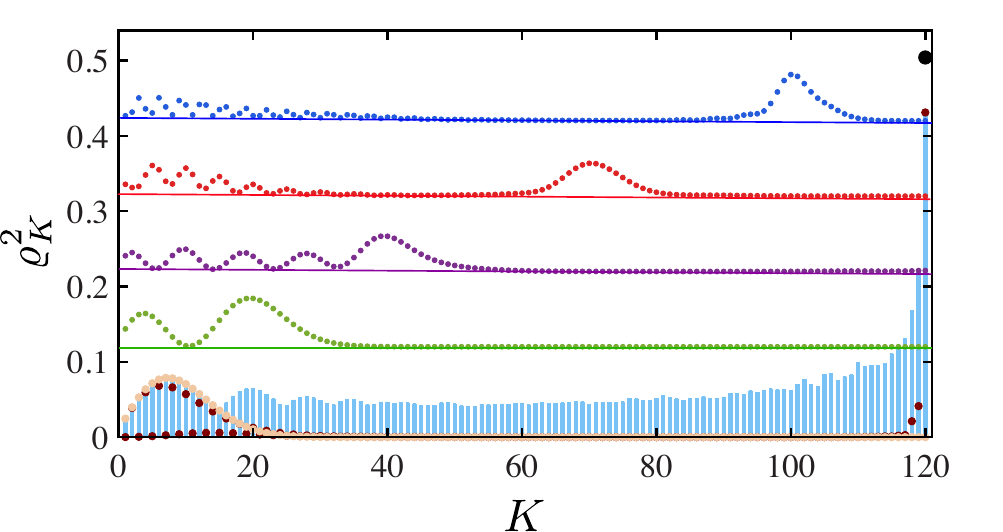}}
	\caption{Maximum multipole length for each order $K$, selected from randomly generated states (blue bars). {Different states maximize the multipoles for different values of $K$, so the} dotted curves represent the multipoles  of the states maximizing a certain region, except for the orange dots, which denote the distribution of the coherent state. The black dot above the last multipole is the value corresponding to the NOON state. Note that the distribution for the coherent state and the state maximizing the last multipole are shown in front of the maximum values, and the rest are vertically displaced for clarity.}
	\label{fig:maxMultipole}
\end{figure}

The results are shown in Fig.~\ref{fig:maxMultipole} for $S=60$. Unexpectedly, it appears that the coherent state {is the unique state with the maximal multipole $\varrho_{K}^2$ for all values of $K$} up to a certain point. After this, another state maximizes {the $K$th multipole for} a different region {of larger $K$ values} and so on. The multipolar distribution of these states is bell shaped near the multipole it maximizes and exhibits a wiggling behavior at the first multipoles. We can also observe how, as $K$ increases, this wiggling evolves to a distribution very similar to that of the NOON state. This supports the hypothesis that the NOON state is the one with the highest contribution to its final multipole {and allows a discrete transition between states maximizing low-order multipoles (coherent) and those maximizing high-order ones (NOON).}.

The King of Quantumness shows the absence of lower-order multipoles, which is the origin of the isotropic behavior of their higher-order fluctuations. However,  the significance of the strength of the first nonzero multipole  is not clear.

We can also gain some general intuition from our calculations. If one has a constellation and makes a change, such as adding, removing, or moving a star, what happens to the multipoles? Upon the addition of a new star $\zeta_{\mathrm{new}}$ to the constellation, for example, the elementary symmetric polynomials change as $e_k \mapsto \zeta_{\mathrm{new}} e_{k-1}+e_k$ for $k>0$ and the rest of the calculations proceed as above. Supposing further that the original constellation corresponded to a coherent state at $z=0$, the addition of a star at $\zeta_{\mathrm{new}}$ makes the multipoles transform to
\begin{align}
    \varrho_{K}^2 = c(K,S) \left [  4 S^2+4 S |\zeta_{\mathrm{new}}|^2+\frac{ \left(K^2+K-2 S\right)^2}{4 S^2}|\zeta_{\mathrm{new}}|^4 \right ] \, ,
\end{align} with 
\begin{equation}
c(K,S)=\frac{(2K+1)(2S-1)!^2|\psi_S|^4}{(2S-K)!(2S+K+1)!} \, .
\end{equation}

\section{Discussion}
\label{sec:fur}

We explore in this section some nontrivial consequences of our method. A first direct result is to compute Stokes operators~\cite{Luis:2000ys} from the constellation. Choosing the $S_z$ operator, for example, all of its moments may now be computed  as
\begin{equation}
    \langle S_z^n\rangle = |\mathfrak{f}_{2S}(\bm{\zeta})|^2  \sum_{m=0}^{2S}
    (m-S)^n \binom{2S}{m}^{-1} |e_{2S-m}(\bm{\zeta})|^2;
\end{equation} 
all other moments may be found by rigidly rotating the constellation to highlight any other operator axis. Knowledge of the Stokes vector $\mathbf{S} = (\langle S_x \rangle, \langle S_y \rangle,\langle S_z\rangle)^{\top}$ (the superscript $\top$ denoting the transpose) and $2S-1$ of the stars provides an overcomplete set of three equations for finding the location of the remaining star. With two stars locations unknown, the covariances between the Stokes parameters provide sufficient information to determine the four unknown angular coordinate parameters, and so on for higher-order moments and determining the unknown locations of more stars. The inversion process involves nonlinear functions and tends to require numerical solutions.

The results may also be used to make contact with spherical $t$-designs, which are sets of points on the sphere that may be used {for} averaging polynomial functions over the entire sphere~\cite{Delsarte:1977dn,Hardin:1992mi}. For a set of points to be a one-design, their vectors must sum to the zero vector. In terms of the stereographic projection, this constrains the points to obey
\begin{equation}
 \sum_{j} \frac{\zeta_{j}}{1 + |\zeta_{j}|^2} = 0 \, ,  \qquad 
\sum_{j} \frac{1-|\zeta_{j}|^2}{1+ |\zeta_{j}|^2} = 0 \, ,  
\end{equation} 
{More connections to designs with $t>1$ are the subject of future study.}

We can further explore the significance of the elementary symmetric polynomials, as they have not yet been used to our knowledge for studying our quantum states. Consider that a Majorana constellation can always be rotated such that one of the stars is at the north pole with $\theta=0$. This makes the highest-order polynomial $e_{2S}$ vanish. Then, the vanishing of the next polynomial $e_{2S-1}$ implies that a second star is also at the north pole. With each subsequent polynomial that vanishes, another star is added to the north pole, up to the condition that $e_j=0\,\forall j>0$ implies that all $2S$ stars are at the north pole and the state is a spin coherent state. The maximal degeneracy of any star in the constellation is then equal to the maximum number of consecutive highest-order symmetric polynomials that vanish, where the latter is maximized over rigid rotations of the stars that perform a known transformation of the roots.

Specifying which polynomials $e_j ( \bm{\zeta}) $ vanish significantly constrains a state. In quantifying quantumness, the vanishing of low-order multipoles is a sign of nonclassicality, while here the vanishing of all higher-order polynomials seems to imply an increase in classicality. We are led to consider an interesting case: what if all of the \textit{lower}-order polynomials vanish, such that the state retains $e_{2S}\neq 0$ while $e_j( \bm{\zeta}) =0\,\forall 0<j<2S$? This includes, for example, the NOON states, which are considered highly quantum by many measures. Since NOON states have all of their stars equally spaced around the equator, the roots all obey $|\zeta_j|=|\zeta_k|$ and are directly equal to roots of unity $\zeta_j = \exp(i \pi  j /S)$. Summing all $2S$ such roots gives $e_1 ( \bm{\zeta})=\sum_{j=1}^{2S}\exp(i \pi  j/S)=0$; summing all $\binom{2S}{2}$ products of such roots gives
\begin{equation}
   e_2( \bm{\zeta}) =\sum_{j<k}^{2S} \exp [ i \pi   (j+k)/S ] = \tfrac{1}{2}\sum_{j\neq k}^{2S} \exp [i \pi (j+k)/S ] = 0 \, ,
\end{equation} 
and so on for all higher-order $e_j( \bm{\zeta})$ other than 
\begin{equation}
e_{2S} ( \bm{\zeta}) =\exp \left ( i \pi \sum_{j=1}^{2S} \frac{j}{S} \right ) =(-1)^{2S+1} \neq 0.
\end{equation}

\begin{figure}
    \centering
    \includegraphics[width=\columnwidth]{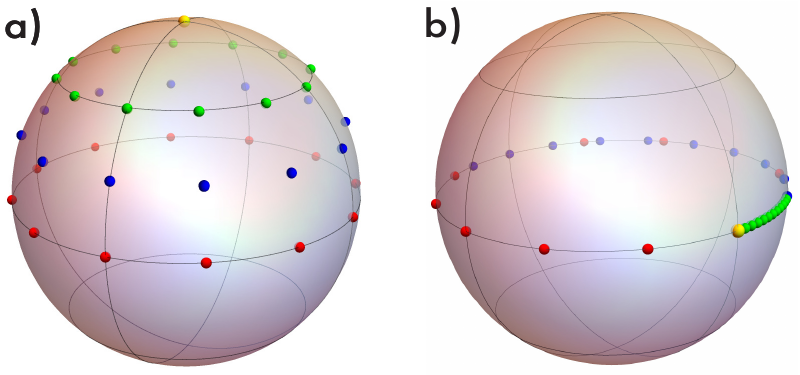}
    \caption{Two visualizations of how a Majorana constellation can smoothly evolve between classical (coherent) and quantum (NOON) states. On the left, the stars are always at equal polar angles to each other and are equally spaced from each other, while the circle they are spaced on transitions between the north pole (coherent) and the equator (NOON). On the right, the stars spread or coalesce around the equator, with equal azimuthal angles between each star and the subsequent star; the final star goes around the equator to come closer to (NOON) or farther from (coherent) wrapping around to reach the initial star from the other side.}
    \label{fig:transition}
\end{figure}

We can visually and mathematically explore the transition between NOON states and coherent states by the migration of stars from the equator to the north pole and by the vanishing of $e_{2S} (\bm{\zeta})$, respectively. Consider the transition as sketched in Fig.~\ref{fig:transition} a), where the $2S$ points in the constellation are equally spaced around a circle other than the equator, sharing the polar angle $\theta$. The symmetric polynomials take the same form as before, with $e_{j} (\bm{\zeta})$  multiplied by $\tan^{j} (\theta/2)$, so all of the lowest polynomials vanish other than 
\begin{equation}
    e_{2S} = (-1)^{2S+1} \tan^{2S}(\theta/2).
\end{equation} 
This codifies the gradual progression: as the circle on which the stars lie gets smaller and smaller, from a NOON to a coherent state, the one nonzero polynomial (other than $e_0=1$) goes to zero as $\tan^{2S}(\theta/2)$. The magnitude of this highest-order polynomial, after a rigid rotation that makes the rest of the polynomials vanish, directly encodes the quantumness of the state. Note that it is the deviation of the magnitude of $e_{2S}$ from unity that matters: there is a coherent state at $\theta=0$ and at $\theta=\pi$.

We can also obtain exact expressions for the multipoles as a function of $\theta$:
\begin{equation}
	\varrho_{K}^2 =  \varrho^{2}_{K, \mathrm{coh}} 
	\left [ \frac{\tan^{4S}(\theta/2)-1}{\tan^{4S}(\theta/2) + 1} \right ]^{2} \, , \quad
	K= 1, 3, \ldots, 2S-1 \, ,
\end{equation}
whereas $	\varrho_{K}^2 =  \varrho^{2}_{K, \mathrm{coh}}$ for $K=2, 4, \ldots 2S-2$.  This transition from coherent to NOON is very abrupt, even for low values of $S$, due to the power of $4S$ in the tangent. The multipoles look much like a coherent state up until the stars are very close to the equator. This shows the vulnerability of the NOON state to a small displacement of its stars.

As previously argued, the NOON state  maximizes the last multipole. With the help of the previous expressions we can obtain this value:
\begin{equation}
 \varrho_{2S,\mathrm{NOON}}^2 = \left \{ 
 \begin{array}{ll}	
 \frac{1}{2}+ \binom{4S}{2S}^{-1} \qquad \qquad &  S=1,2,\ldots \\
 & \\
 \frac{1}{2}  &  S=\frac{1}{2},\frac{3}{2},\ldots
 \end{array}
 \right .
 \end{equation}
which tends to $1/2$ as $S$ becomes large. Note that this does not contradict the results in Fig.~\ref{Fig:Multcoh}, for the multipoles are normalized therein.

\begin{figure}
    \centering
    \includegraphics[width=0.85\columnwidth]{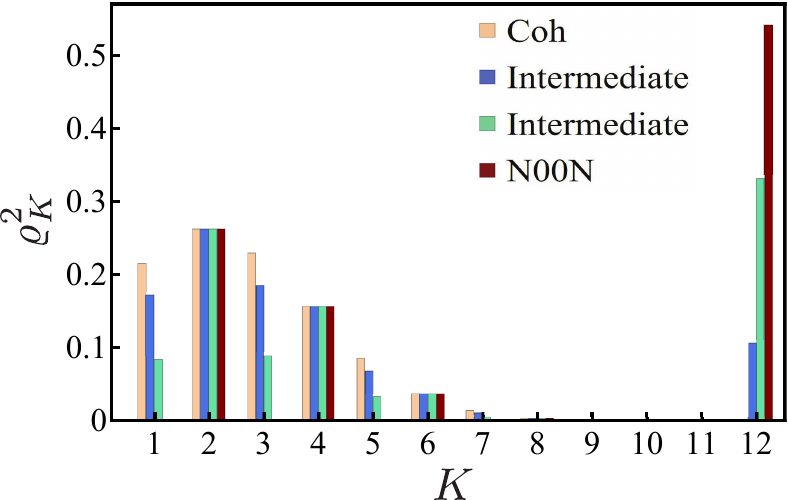}
    \caption{Evolution of the multipole lengths corresponding to case a) in Fig.~\ref{fig:transition}. The position of the stars in the figure is for illustrative purposes only, as the transition occurs when the stars are very close to the equator.}
    \label{fig:multipole transitions}
\end{figure}

We can observe how the multipole lengths change in the aforementioned transition between coherent and NOON states in Fig.~\ref{fig:multipole transitions}. This transition can be enacted by evolution under the highly nontrivial Hamiltonian
\begin{equation}
    H_{\mathrm{NOON}\leftrightarrow\mathrm{coh}}= -i |S,S\rangle\langle S,-S|+ 
    i |S,-S\rangle\langle S,S| 
\end{equation}
that manifestly breaks degeneracies between Majorana stars.

Further consider the alternative transition as in Fig.~\ref{fig:transition} b), where the stars spread around the equator ($\theta=\pi/2$) smoothly between NOON- and coherent-state configurations. Now we consider the coherent state to have $\zeta=1$, such that its polynomials take the form $e_j (\bm{\zeta}) =\binom{2S}{j}$. We choose the spreading rate such that all stars arrive in place at the same time, with $\zeta_j (t)=\exp(i \pi t j/S)$ and $t\in(0,1)$ with $t=0$ and $t=1$ corresponding to coherent and NOON states, respectively. Then
\begin{equation}
    \begin{aligned}
        e_1( \bm{\zeta}(t)) &= \exp \left [ i \pi  \frac{(2 S+1)}{2S} t \right ] \frac{\sin (\pi t)}{\sin \left ( \frac{\pi t}{2S} \right ) } \, ,\\
    e_2 ( \bm{\zeta}(t)) & = \tfrac{1}{2!} [ e_1( \bm{\zeta}(t))^{2} - e_1( \bm{\zeta}(2t)) ] \, ,  \\
    e_3 ( \bm{\zeta}(t)) & = \tfrac{1}{3!} [e_1( \bm{\zeta}(t))^{3} - 3 e_1( \bm{\zeta}(t)) e_1( \bm{\zeta}(2t))+ 2 e_1( \bm{\zeta}(3t))] \, ,
    \end{aligned}
\end{equation} 
and so on. For small $t$, these equal 
\begin{equation}
e_j (\bm{\zeta}(t))= \binom{2S}{j}-\binom{2S}{j+1}\pi^2  \frac{2S+1}{4S^2} t^{2}+\mathcal{O}(t^4) \, .
\end{equation}  
The Hamiltonian required to enact this transition is much more complicated, as the stars all rotate around the $z$-axis but no Hamiltonian of the form $S_z^n$ can do anything other than rotate the NOON state.

Finally, consider the symmetric transition where the stars spread in both directions along the equator, with pairs moving at equal speeds. For example, with $2S$ odd, we can set $\zeta_0=1$, $\zeta_j (t)= \exp(i \pi t j/S)$ for $0<j\leq S-1/2$, and relabel the remainder as $\zeta_{-j}(t)=\exp(- i \pi  t j/S)$ for $0<j\leq S-1/2$. The spacings between all of the stars are the same as when all of the stars travelled in the same direction, merely offset by a relative phase, which symmetrizes the expressions to
\begin{equation}
    \begin{aligned}
        e_1( \bm{\zeta}(t)) &= \frac{\sin (\pi t)}{\sin \left ( \frac{\pi t}{2S} \right ) } \, ,\\
       e_2 ( \bm{\zeta}(t)) & = \tfrac{1}{2!} [ e_1( \bm{\zeta}(t))^{2} - e_1( \bm{\zeta}(2t)) ] \, ,   
    \end{aligned}
 \end{equation}
and so on. We can thus use the symmetric polynomials to observe the transitions between highly quantum and highly classical states, in accordance with the geometric picture of the Majorana constellation.

\section{Concluding remarks}
\label{sec:conc}

The Majorana representation provides a valuable geometric tool to characterize quantum states with SU(2) symmetry. The visualization of quantum states as a constellation in the unit sphere is finding new applications in quantum information science. On the other hand, the multipolar expansion is especially germane to capture the state invariant properties and, in addition, the resulting multipoles can be measured in the laboratory.  

We have extensively explored how to go from Majorana constellations to multipolar distributions and back, illustrating this through a variety of examples. For instance, we have elucidated the changes in the multipolar distribution when the Majorana constellation transitions from a classical (concentrated) to a highly quantum state (spread out). Additionally, we have delved into the consequences of adding new stars to the constellation and examined states with the greatest contribution to the higher-order multipole, among many other aspects. Interestingly, the formalism can be manifestly extended to other symmetries~\cite{Fabre:2023aa}: this is more than an academic curiosity, and work in this direction is ongoing.

\acknowledgments

We are indebted to G. Bj\"{o}rk, P. de la Hoz, M. Grassl, L. Villanueva, and \L. Rudnicki for discussions. We acknowledge financial support from the European Union QUANTERA program (project ApresSF), and the Spanish Researcy Agency (Grant No. PID2021-127781NB-I00). A.Z.G. acknowledges funding from the NSERC PDF program. A.Z.G. acknowledges that the NRC headquarters is located on the traditional unceded territory of the Algonquin Anishinaabe and Mohawk people. 

%

\end{document}